\documentstyle[epsfig,prc,preprint,aps]{revtex}

\begin{document}
\draft
\begin{title}
{Numerical renormalization using dimensional regularization:
a simple test case in the Lippmann-Schwinger equation}
\end{title}

\preprint{NT@UW-99-45}

\author{D.R. Phillips$^{\rm a}$, I. R. Afnan$^{\rm b}$ and 
A.G. Henry-Edwards$^{\rm b}$\\
$^{\rm a}$ Department of Physics, University of Washington, \\ 
Seattle, WA 98195-1560, U.S.A.\\
$^{\rm b}$ Department of Physics, The Flinders University of South Australia,\\ 
G.P.O. Box 2100, Adelaide 5001, Australia}

\date{\today}

\maketitle

\begin{abstract} 
Dimensional regularization is applied to the
Lippmann-Schwinger equation for a separable potential which gives rise
to logarithmic singularities in the Born series. For this potential a
subtraction at a fixed energy can be used to renormalize the amplitude
and produce a finite solution to the integral equation for all
energies. This can be done either algebraically or numerically.  In
the latter case dimensional regularization can be implemented by
solving the integral equation in a lower number of dimensions, fixing
the potential strength, and computing the phase shifts, while
taking the limit as the number of dimensions approaches three. We
demonstrate that these steps can be carried out in a numerically
stable way, and show that the results thereby obtained agree with
those found when the renormalization is performed algebraically to
four significant figures.  
\end{abstract}


\section{Introduction}\label{sec1}

One difficulty for standard treatments of hadronic reactions is that
form factors are introduced at hadronic vertices in order to regulate
integrals which would otherwise be divergent. This procedure reflects
the substructure of hadrons which gives them a finite extent, and
hence a form factor. However, any field theory upon which such
calculations are based will necessarily be non-local. The
implementation of basic field theoretic principles, such as causality
and electromagnetic gauge invariance, is quite involved in such field
theories~\cite{CP54,CP53}. Methods which impose gauge invariance on an
amplitude containing hadronic form factors have been
formulated,\cite{Oh89,AA95,Ha97} but they are intrinsically
non-unique, since they constrain only the longitudinal part of the
photon's coupling to the hadronic system.

Some of these difficulties can be resolved by using dimensional
regularization (DR)\cite{HV72,BG72,Le75,IZ84} to render divergent
integrals finite. This is the method of choice for dealing with the
infinities which arise in perturbative field theoretic calculations.
However, there are very few studies of the application of dimensional
regularization to integral equations. One notable exception is the
recent application of DR to a Schwinger-Dyson equation in quenched QED
by Schreiber {\it et al.}\cite{ASW98}. In this paper we adapt their
method and apply it to the Lippmann-Schwinger (LS) equation.

We consider one member of the class of potentials in which one
subtraction at a fixed energy results in a finite integral equation at
all energies. The potential we choose is separable and so provides a toy
problem where the renormalized amplitude can be derived algebraically
in a straightforward way. The amplitude thereby obtained can then be
compared with that found by the numerical solution of the
dimensionally-regulated integral equation. We will show that with
careful numerical work the phase shifts obtained via these two
different methods are in excellent agreement with each other. This
proves that many of the difficulties associated with implementing
dimensional regularization numerically can be overcome, and hence that
such ``numerical dimensional regularization and renormalization"
facilitates the extraction of finite phase shifts from a particular
class of divergent potentials.

In Sec.~\ref{sec2} we outline the general framework for implementing
dimensional regularization and renormalization techniques numerically
in the LS equation. We show how the LS equation can be analytically
continued to lower numbers of dimensions, $D=3-\epsilon$, thus
ameliorating the divergences which appear in three dimensions.  The
application of a renormalization condition at one energy in
$3-\epsilon$ dimensions can then lead to an amplitude which is finite
at all energies in this lower number of dimensions, and thence to an
amplitude which is well-defined as the limit $\epsilon \rightarrow 0$
is taken. In Sec.~\ref{sec3} the special case of a separable
potential, which leads to a divergent amplitude as $\epsilon
\rightarrow 0$ is considered. We obtain the amplitude for this
potential for $\epsilon > 0$. Renormalizing the amplitude by demanding
that the binding energy of the deuteron be reproduced leads to a
divergence in the inverse of the ``bare" coupling which appears in the original
separable potential. We show that in DR this divergence goes as
$1/\epsilon$.  We then observe that for this 
particular potential, it is not necessary to use DR, since the
problem can be subtractively renormalized. That is to say, a
subtraction at one energy, which can be carried out algebraically,
renders the amplitude finite at all energies. 

However, it is not generally true that such algebraic renormalization
of an integral equation is possible. Therefore, in Section~\ref{sec4}
the general renormalization procedure of Section~\ref{sec2} is applied
directly to the LS equation in $3 - \epsilon$ dimensions. In this
approach no subtraction is carried out, and no assumption is made
about the form of the potential. First, the LS equation is solved
numerically in $D=3 - \epsilon$ dimensions.  Then, demanding a
specific value of the amplitude at one energy implies a value of the
bare coupling in $D$ dimensions. The amplitude can then be computed
numerically for this value of the coupling at arbitrary energy and
physical quantities, such as phase shifts, extracted in $D$
dimensions. This must be done over a range of values $\epsilon$ and
then the limit $\epsilon \rightarrow 0$ can be taken. Finally, in
Sec.~\ref{sec5} we present some concluding remarks.


\section{Theoretical Analysis}\label{sec2}

Dimensional regularization is often used to identify the infinities
encountered in perturbative calculations~\cite{HV72,BG72,Le75,IZ84}.
The procedure involves replacing integrals that are infinite in $n$
dimensions by the same integrals in a lower dimension $n-\epsilon$,
where they have a finite result. The function defining this integral,
say $I(\epsilon)$, can then be analytically continued back to $n$
dimensions.  In this way the logarithmic singularities of $I$ can be
isolated and then renormalized by adjusting the bare parameters of the
theory.  This is the method of choice for dealing with the infinities
which arise in perturbative quantum field theory calculations, but it
has not been widely applied to non-perturbative integral equations
that contain divergences.  It was used in the Lippmann-Schwinger
equation for a divergent potential in Refs.~\cite{Ka96,Ph97}, but in
the cases considered there the amplitude can be derived analytically.
In the case of a general divergent potential, this will not be
possible, and DR must be implemented numerically.  Recently, Schreiber
{\it et al.}~\cite{ASW98}  have demonstrated the feasibility of such a procedure,
successfully employing DR in the numerical solution of a divergent
four-dimensional Schwinger-Dyson equation for the electron mass in
quenched QED.  In this section we explain how to adapt the methods of
Schreiber {\it et al.} for use in the three-dimensional
Lippmann-Schwinger equation.

Given a potential $V$ the momentum-space LS equation for the
scattering amplitude $T$ for two particles of mass $m$ is:

\begin{equation} 
T({\bf k},{\bf k}';E) = V({\bf k},{\bf k}') + \int
d^3 k'' \, V({\bf k},{\bf k}'') \, \frac{1}{E - k''^2/m + i \eta} \,
T({\bf k}'',{\bf k};E) \ , 
\label{eq:2.0.1} 
\end{equation} 
where $\eta$ is a positive infinitesimal.  To dimensionally-regulate
this equation we must continue all quantities in it into $3 -
\epsilon$ dimensions. If the potential $V$ is central, and
depends only on the angle between ${\bf k}$ and ${\bf k}'$, then the
operator $L^2$ commutes with the Hamiltonian. It follows that $V$ and
$T$ can be expanded using the eigenfunctions of the rotation operator
in $3 - \epsilon$ dimensions, thus reducing Eq.~(\ref{eq:2.0.1}) to a
one-dimensional integral equation. This expansion can be quite
involved (see, for instance, the analogous expansion in $4 - \epsilon$
dimensions of Ref.~\cite{ASW98}).

Here we follow a slightly different approach. Although the
Lippmann-Schwinger equation is nominally a three-dimensional integral
equation, in most instances it is written and solved as a
one-dimensional integral equation for a given partial wave $\ell$.  In
other words, making the expansion:

\begin{equation} 
T({\bf k},{\bf k}';E) =\frac{1}{4 \pi} \sum_\ell (2
\ell + 1) T^\ell(k,k';E) P_l(\hat{k} \cdot \hat{k}') \ , 
\label{eq:2.0.2}
\end{equation}
and an equivalent expansion for $V$, leads, as is well-known, to the
one-dimensional integral equation:

\begin{equation} 
T^\ell(k,k';E) = V^\ell(k,k') + \int\limits_0^\infty\,dk''\,k''^2\
\frac{V^\ell(k,k'')}{E-k''^2/m+i\eta}\,T^\ell(k'',k';E) \ . 
\label{eq:2.0.3}
\end{equation}

To dimensionally-regulate {\it this} equation is much easier, since
the angular part of the problem has already been dealt with in the
original number of dimensions, in this case, three.  It only involves
replacing the measure appropriate to three dimensions, by one for
$(3-\epsilon)$ dimensions, i.e.  

\begin{equation} 
T^\ell_\epsilon(k,k';E) = V^\ell(k,k') +
\int\limits_0^\infty\,dk''\, k''^{(2-\epsilon)}\
\frac{V^\ell(k,k'')}{E-k''^2/m + i \eta} \,T^\ell_\epsilon(k'',k';E) \ .
\label{eq:2.0.4} 
\end{equation}

It is worth pointing out here that if the partial-wave expansion
(\ref{eq:2.0.2}) is carried out in $3-\epsilon$ dimensions, rather
than in three dimensions, then the potential $V^\ell$ ends up
depending on $\epsilon$ as well as $\ell$. (See, for instance,
Ref.~\cite{ASW98}.)  However, if we are only concerned with isolating
the divergences of the integral equation as $\epsilon \rightarrow 0$,
then we are free to take the limit as $\epsilon \rightarrow 0$ first
in $V^\ell$, since that limit is non-singular. Thus, choosing to
dimensionally regularize the partial-wave-expanded equation, rather
than the full three-dimensional equation, may be unfamiliar to
those used to DR in perturbative settings, but it is justified.

Now consider a potential $V^\ell(k',k)$ which leads to divergences 
in the Born series for Eq.~(\ref{eq:2.0.3}).  The integral equation
(\ref{eq:2.0.4}) will yield finite answers for the terms in that
series, provided that $\epsilon$ is taken large
enough. Equation (\ref{eq:2.0.4}) has been regularized. The next step is to
renormalize it. To that end we pick one of the parameters of the
potential, e.g. its strength, which we denote by $\lambda$, and
regard it as a function of $\epsilon$, $\lambda=\lambda_\epsilon$. A
renormalization condition is then chosen. This can be represented as
the demand that the amplitude, $T^\ell_\epsilon(k,k';E)$, have some
specified value at a particular kinematic point. For instance, we
might demand that the amplitude has a pole at $E=-B$, so that a bound
state exists at that energy. Such a renormalization condition
implicitly defines the function $\lambda_\epsilon$. The amplitude
$T^\ell_\epsilon(k',k;E)$ can then be computed at the on-shell point:
$k'^2=k^2=m E$, for a variety of energies, with this particular choice
of $\lambda_\epsilon$. This will certainly give finite results for
$\epsilon > 0$, provided that $\epsilon$ is big enough. Furthermore,
if it is enough to renormalize only the strength of the potential,
such a calculation will produce finite values for the on-shell
amplitude and hence for the phase shift even in the limit $\epsilon
\rightarrow 0$. This ``numerical renormalization" program has been
carried out before in the Lippmann-Schwinger equation with other
regulators (see, e.g.~\cite{Sc97,Le97,Ge98,Gh98}), but here we will
implement it for DR.

In the next section we illustrate how this numerical renormalization
proceeds in the case of a simple separable potential. We show that the
strength of the potential can always be adjusted so that the potential
supports a bound state with energy $E=-B$. This strength cancels the
divergence in the Born series integrals, and so a finite scattering
amplitude can be extracted.


\section{The separable potential and algebraic renormalization}\label{sec3}

Consider a separable potential with a form factor that does not
guarantee the convergence of the integrals needed to determine the
scattering amplitude. An example of such a potential is (from this
point on we drop the angular momentum label $\ell$)

\begin{equation}
V(k,k') = g(k)\,\lambda\,g(k') \quad\mbox{with}\quad g(k) =
\frac{1}{(k^2 + \beta^2)^{1/4}} \ .
\label{eq:2.1.1} 
\end{equation} 
The corresponding off-shell $T$-matrix is given by 
\begin{equation}
T(k,k';E^+) = g(k)\,\tau(E^+)\,g(k') \ ,
\label{eq:2.1.2} 
\end{equation}
with 
\begin{equation} \left[\tau(E)\right]^{-1} = \lambda^{-1} - I(E) \ ,
\label{eq:2.1.3} 
\end{equation} 
where the integral $I(E)$ is
\begin{equation} I(E) = m\,\int\limits^\infty_0\, dk\,k^2\
       \frac{[g(k)]^2}{mE - k^2 + i\eta}       \ .
\label{eq:2.1.4}
\end{equation}
Since $g(k)\rightarrow k^{-1/2}$ as $k\rightarrow \infty$, the integrand in
Eq.~(\ref{eq:2.1.4}) goes as $k^{-1}$ for $k\rightarrow\infty$, and the
integral $I(E)$  has a logarithmic divergence.

Of course, if we had inserted the potential (\ref{eq:2.0.2}) in the
dimensionally-regulated Lippmann-Schwinger equation, then the only
thing that would have changed in the above analysis is that equations
(\ref{eq:2.1.3}) and (\ref{eq:2.1.4}) would have become:

\begin{eqnarray} 
\left[\tau_\epsilon(E)\right]^{-1} &=& \lambda_\epsilon^{-1}
   - I_\epsilon(E) 
\label{eq:2.1.5a}\\
I_\epsilon(E)&=& m \, \int \limits^\infty_0 \,  dk  \, k^{2 - \epsilon}\
       \frac{[g(k)]^2}{mE - k^2 + i\eta}       \ .
\label{eq:2.1.5b}
\end{eqnarray}

This integral is now finite for any $\epsilon$ greater than zero.  The
logarithmically divergent part of $I_\epsilon$ will stand revealed as
a $1/\epsilon$ pole.  Since $\lambda_\epsilon$ is a function of
$\epsilon$ here this pole can be cancelled by making an appropriate
choice of $\lambda_\epsilon$.  Specifically, here we achieve this
renormalization by demanding that one observable be reproduced,
i.e. that the $^3$S$_1$ $NN$ channel have a bound state at energy
$E_B= -2.2246$~MeV. This is done so that our interaction can be
regarded as a toy model of the $NN$ potential in the deuteron
channel. This model is extremely simple, but it has the only two
features we are concerned with here: it leads to logarithmic
divergences in the Born series, and it has a shallow bound state. The
renormalization condition which fixes the position of this bound state
may be expressed as follows:

\begin{equation} 
[\tau_\epsilon(E)]^{-1} \equiv \lambda_\epsilon^{-1} -
I_\epsilon(E) \rightarrow 0 \quad\mbox{as}\quad E\rightarrow E_B = -B
<0 \ .  
\label{eq:2.1.6}
\end{equation}

To see how this removes the divergent part of $I_\epsilon(E)$, we
write the form factor $g(k)$ as
\begin{equation}
g(k) \equiv \frac{1}{\sqrt{k}}\ h(k)   \quad\mbox{with}\quad
h(k)\rightarrow 1\quad \mbox{as}\quad k\rightarrow\infty \ .  
\label{eq:2.1.7}
\end{equation}
Our singular integral can then be written as
\begin{equation}
I_\epsilon(E) = - m\ \int\limits^\infty_0\ dk\ k^{(1-\epsilon)}\ 
              \frac{[h(k)]^2}{k^2 - mE - i\eta} \ .
\label{eq:2.1.8}
\end{equation}
We now define an integral $\tilde{I}_\epsilon(E)$ that retains the 
singular part of $I_\epsilon(E)$ in the limit as $\epsilon\rightarrow 0$:
\begin{equation}
\tilde{I}_\epsilon(E) \equiv - m\ \int\limits^\infty_0\ dk \ 
        \frac{k^{(1-\epsilon)}}{k^2 - mE -i\eta}\ .
\label{eq:2.1.9}
\end{equation}
This allows us to write $I_\epsilon(E)$ in terms of a subtraction as
\begin{eqnarray}
I_\epsilon(E) &=& - m\ \int\limits^\infty_0\ dk\ k^{(1-\epsilon)}\ 
\left\{\frac{[h(k)]^2}{k^2-mE-i\eta} - \frac{1}{k^2+mB}\right\}
+ \tilde{I}_\epsilon(-B)                          \nonumber \\
 &\equiv& J_\epsilon(E) + \tilde{I}_\epsilon(-B)\ , 
\label{eq:2.1.10}
\end{eqnarray}
where the integral $J_\epsilon(E)$ is well defined in the limit as
$\epsilon\rightarrow 0$. Making use of the identity:
\begin{equation}
\int\limits^\infty_0\ dr\,\frac{r^\beta}{(r^2 + A)^\alpha} = 
 \frac{1}{2}\ \frac{\Gamma\left(\frac{\beta+1}{2}\right)\
 \Gamma\left(\alpha - \frac{\beta+1}{2}\right)}{A^{\alpha -
 (\beta+1)/2}\ \Gamma(\alpha)}  \ ,                   
\label{eq:2.1.11}
\end{equation}
we have
\begin{equation}
\tilde{I}_\epsilon(-B) = - \frac{m}{\epsilon}\
\frac{\Gamma\left(1-\epsilon/2\right)\ 
  \Gamma\left(1+\epsilon/2\right)}{(mB)^{\epsilon/2}} \rightarrow -
\frac{m}{\epsilon} \quad \mbox{as}\ \epsilon\rightarrow 0 \ .
\label{eq:2.1.12}
\end{equation} 
This isolates the singularity of $I_\epsilon(E)$ in the limit
$\epsilon\rightarrow 0$. We now need to renormalize the strength of the
potential $\lambda_\epsilon$ in order to get a finite amplitude.

To do this we write the condition for a bound state (\ref{eq:2.1.6}) as
\begin{equation}
\lambda_\epsilon^{-1} = I_\epsilon(-B)  
\label{eq:2.1.13}
\end{equation}
i.e.
\begin{equation}
\lambda_\epsilon^{-1} = - \frac{m}{\epsilon}\ 
\frac{\Gamma(1-\epsilon/2)\ \Gamma(1+\epsilon/2)}{(mB)^{\epsilon/2}}
 + J_\epsilon(-B) \ , 
\label{eq:2.1.14} 
\end{equation} 
which makes it clear that $\lambda_\epsilon^{-1}$ has both a divergent
part, proportional to $1/\epsilon$ in DR, and other, finite, parts, that
are dependent upon the particular bound-state energy we are trying to
reproduce. Equation (\ref{eq:2.1.14}) gives $\lambda_\epsilon$ for 
a given $\epsilon$ and from that we can then calculate 
$\tau_\epsilon(E)$ and the $T$-matrix.

Alternatively, we can use Eq.~(\ref{eq:2.1.13}) for 
$\lambda_\epsilon$ to write:
\begin{equation}
\tau_\epsilon^{-1}(E)=I_\epsilon(-B) - I_\epsilon(E)\ .
\label{eq:2.1.15}
\end{equation}
Inserting Eq.~(\ref{eq:2.1.5b}) for $I_\epsilon$ and combining the 
integrands we see that:
\begin{equation}
\tau_\epsilon^{-1}(E)=(E + B) \int\limits^\infty_0\ dk\ 
k^{(1-\epsilon)} \ \frac{[h(k)]^2}{(k^2/m - E - i \eta)(B + k^2/m)} \ .
\label{eq:2.1.16}
\end{equation}
which implies that:
\begin{equation}
T_\epsilon(k',k;E)=g(k') \, \frac{S_\epsilon(E)}{E+B} \, g(k), 
\label{eq:2.1.17}
\end{equation}
where:
\begin{equation}
S_\epsilon^{-1}(E)=\int\limits^\infty_0\ dk\ k^{(1-\epsilon)} \ 
              \frac{[h(k)]^2}{(k^2/m - E - i \eta)(B + k^2/m)} 
\label{eq:2.1.18} 
\end{equation} 
is a finite quantity. Since $S_\epsilon(E)$ is finite, the limit as
$\epsilon$ goes to zero can be taken smoothly, and we never even need
to consider any quantity other than $S_0(E)$, when the regulator has
been removed altogether. It is now trivial to obtain
phase shifts from Eqs.~(\ref{eq:2.1.17}) and (\ref{eq:2.1.18}). Their
relationship to the on-shell amplitude is just:
\begin{equation}
\tan(\delta)=\frac{\mbox{Imag }T(E;k_0,k_0)}
                  {\mbox{Real }T(E;k_0,k_0)},
\label{eq:2.1.19}
\end{equation}
where $k_0^2=mE$ is the on-shell momentum. 


\section{Numerical renormalization}\label{sec4}

The question motivating this study is whether numerical techniques can
be found that give the amplitude $T_\epsilon(k',k;E)$ numerically when the LS
equation is solved in less than three dimensions. These techniques
must be stable enough to allow the limit $\epsilon \rightarrow 0$ to
be taken. For the simple potential (\ref{eq:2.1.1}) the result of any
such procedure must be given by Eqs.~(\ref{eq:2.1.17}) and
(\ref{eq:2.1.18}), where an ``algebraic renormalization" making explicit use
of the potential's separability has been carried out. 

In this section we implement an alternative strategy to this algebraic
renormalization, that of Section~\ref{sec2}, which we call ``numerical
renormalization". We solve the partial-wave LS equation in
$(3-\epsilon)$ dimensions, Eq.~(\ref{eq:2.0.2}), directly. The strength
$\lambda_\epsilon$ is adjusted to give the correct bound-state energy,
i.e. so that $T_\epsilon(k,k';E)$ has a pole at $E=-B$. Once this is done
we calculate the phase shifts as $\epsilon\rightarrow 0$. Here the
dimensionally-regularized integral equation is being used to
renormalize the strength of the potential. We do not have to
explicitly carry through the subtractive, algebraic renormalization
discussed in Sec.~\ref{sec3}, and the dimensionally-regularized LS
equation should give phase shifts as a function of $\epsilon$ which
are well-behaved in the limit $\epsilon \rightarrow 0$.

The two signals we look for to determine the success of our numerical
implementation of DR for our divergent integral equation are:
\begin{enumerate} 
\item $\lambda_\epsilon$ must have the behavior
(\ref{eq:2.1.14}) in the limit $\epsilon \rightarrow 0$;
\item The results for phase shifts from numerical renormalization must be
stable in the limit $\epsilon \rightarrow 0$.
\end{enumerate}

These two tests should be applicable to {\it any}
dimensionally-regularized integral equation. However, in the case
under investigation here the results of Eqs.~(\ref{eq:2.1.17}) and
(\ref{eq:2.1.18}) provide an additional check on the accuracy of this
numerical renormalization using dimensional regularization.

To perform such tests, we need to solve the integral equation for each
value of $\epsilon$ for both the bound state problem and the
scattering problem with considerable accuracy. To achieve this
accuracy, we first split the integral in the Lippmann-Schwinger
equation arbitrarily into a piece from zero to some $k_m$ 
and a piece from $k_m$ to infinity. Then, on the
interval $[k_m,\infty)$ we make a change of variables to:

\begin{equation} 
t=\epsilon \log\left(\frac{k}{k_m}\right), 
\label{eq:changequads}
\end{equation} 
thereby enabling us to employ a logarithmic mesh which ensures the
correct numerical integration when we obtain $1/\lambda_\epsilon$.  
The variable transformation (\ref{eq:changequads}) introduces a factor
$e^{-t}$ into the integrand, and so Gauss-Laguerre quadratures are
chosen for the integration from $k_m$ to infinity, since they
naturally build in this factor.

\begin{figure}[h]
\vskip 0.5 cm
\centerline{\epsfig{figure=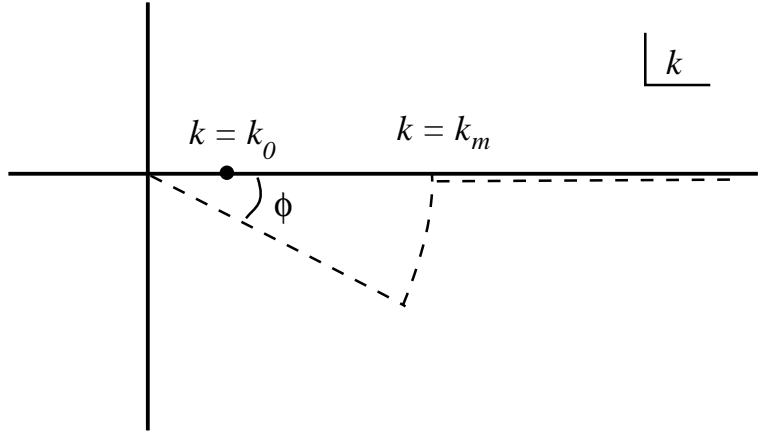, width=10.0cm}}
\caption{The rotated contour of integration. Here $k_0$ is
the on-shell momentum}\label{fig.1} 
\end{figure}

Meanwhile, ordinary Gauss-Legendre quadratures are employed for the 
integration on $[0,k_m]$. This was done in two ways when
we performed ``numerical renormalization":

\begin{enumerate}
\item We deformed the contour of integration from the real
$k$-axis to that depicted in Fig.~\ref{fig.1}. Along the contour from
zero to $k_m e^{-i\phi}$ we divide the interval into two parts with
$n_0$ quadratures for $0\rightarrow 2k_0e^{-i\phi}$ and $n_1$
quadratures for $2k_0e^{-i\phi}\rightarrow k_me^{-i\phi}$. Finally, we
take $n_2$ quadratures for the part of the contour that returns us to
the real axis, and $n_3$ for the interval $[k_m,\infty)$.  In
this way we can optimize the four different regions of the integration
independently. For the determination of $\lambda$ by the requirement
that the potential supports a bound state, we take $n_0=0$, i.e.
the contour of integration corresponds to taking $k_0=0$. We have
optimized the number of quadratures on each of the intervals of
integration as well as the angle $\phi$ and the point at which the
contour returns to the real $k$ axis, $k_m$. We have found it
necessary to take $n_0=16, n_1=80, n_2=10, n_3=15, \phi=0.7$ and
$k_m=50$~fm$^{-1}$. We could have used a smaller number of quadratures
on each interval, but to establish that DR is valid for the LS
equation and for the two methods to give {\it identical} results, we
have not economized on the number of quadratures.

\item Instead of solving for the T-matrix we solved for the K-matrix
and used the relationship:

\begin{equation} T^\ell(E^+)=K^\ell(E)-\frac{1}{2}\,i \pi m k_0\,
K^\ell(E)\, T^\ell(E^+), \end{equation} to relate the two on-shell,
and extract the phase shifts. The K-matrix is, of course, a purely
real quantity, but the integral equation defining it has a
principal-value singularity at $k=k_0$, the on-shell point. To deal
with this we place $n_0$ Gauss-Legendre quadratures, distributed
symmetrically about $k=k_0$, on the interval $[0, 2 k_0]$. We then
also place $n_1$ Gauss-Legendre quadratures on $[2 k_0,k_m]$ and $n_3$
Gauss-Laguerre quadratures on $[k_m,\infty)$.  Here we found it
sufficient to take $n_0=16$, $n_1=60$, $n_3=15$, and
$k_m=50$~fm$^{-1}$.  
\end{enumerate} 
Each of these methods for solving the integral equation is accurate to
four significant figures, provided that $\epsilon > 10^{-6}$. For
very small $\epsilon$ only the second mesh produces results which are
this stable.

Note that only the first integration technique was employed to do the
integration in Eqs.~(\ref{eq:2.1.17}) and (\ref{eq:2.1.18}), but there
the precise details are less important, since the integral in question
is finite. 

\begin{figure}[t] \centerline{\epsfig{figure=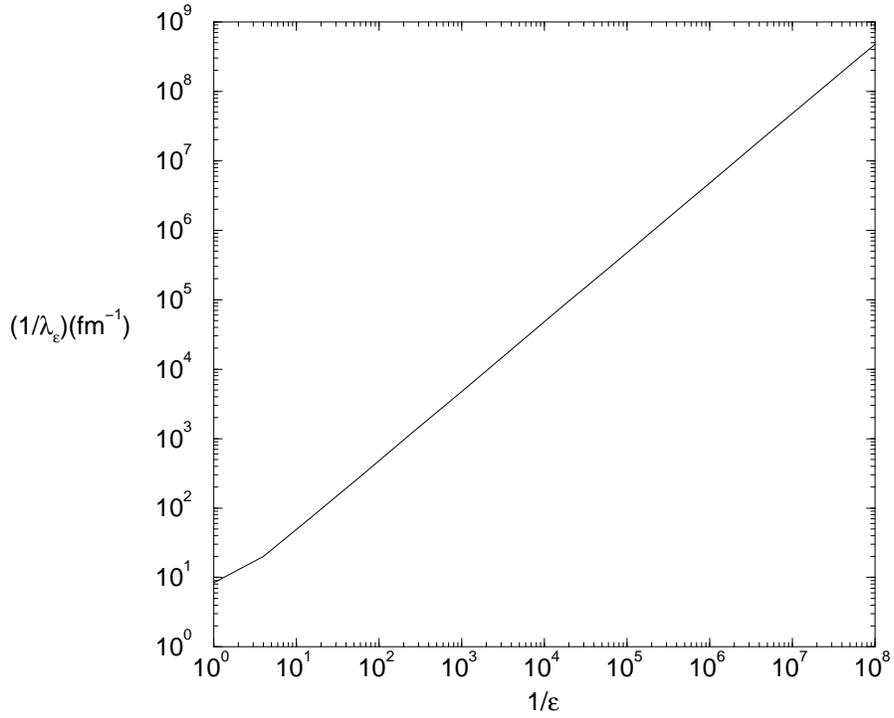,
width=12.0cm}} \caption{The behaviour of the inverse of the bare
coupling $\lambda$ as a function of $1/\epsilon$, as obtained
using the dimensionally-regularized Lippmann-Schwinger equation}
\label{fig.lambda} 
\end{figure}

The first check that the dimensional regularization of this integral
equation is being done correctly is to see that the bare coupling
$\lambda$ extracted by imposing the condition that there be a bound
state at $E=-B$ on Eq.~(\ref{eq:2.0.4}) does behave according to
Eq.~(\ref{eq:2.1.14}).  In Fig.~\ref{fig.lambda} we plot the function
$1/\lambda_\epsilon$ versus $1/\epsilon$ that is found when this
condition is imposed on Eq.~(\ref{eq:2.0.4}) in the case of the
potential (\ref{eq:2.1.1}). We chose $B=-2.225$ MeV, and
$\beta=1.4489$ fm$^{-1}$.  The slope of the curve, is indeed, $m$, the
nucleon mass, as per Eq.~(\ref{eq:2.1.14}).

\begin{figure}[h] 
\centerline{\epsfig{figure=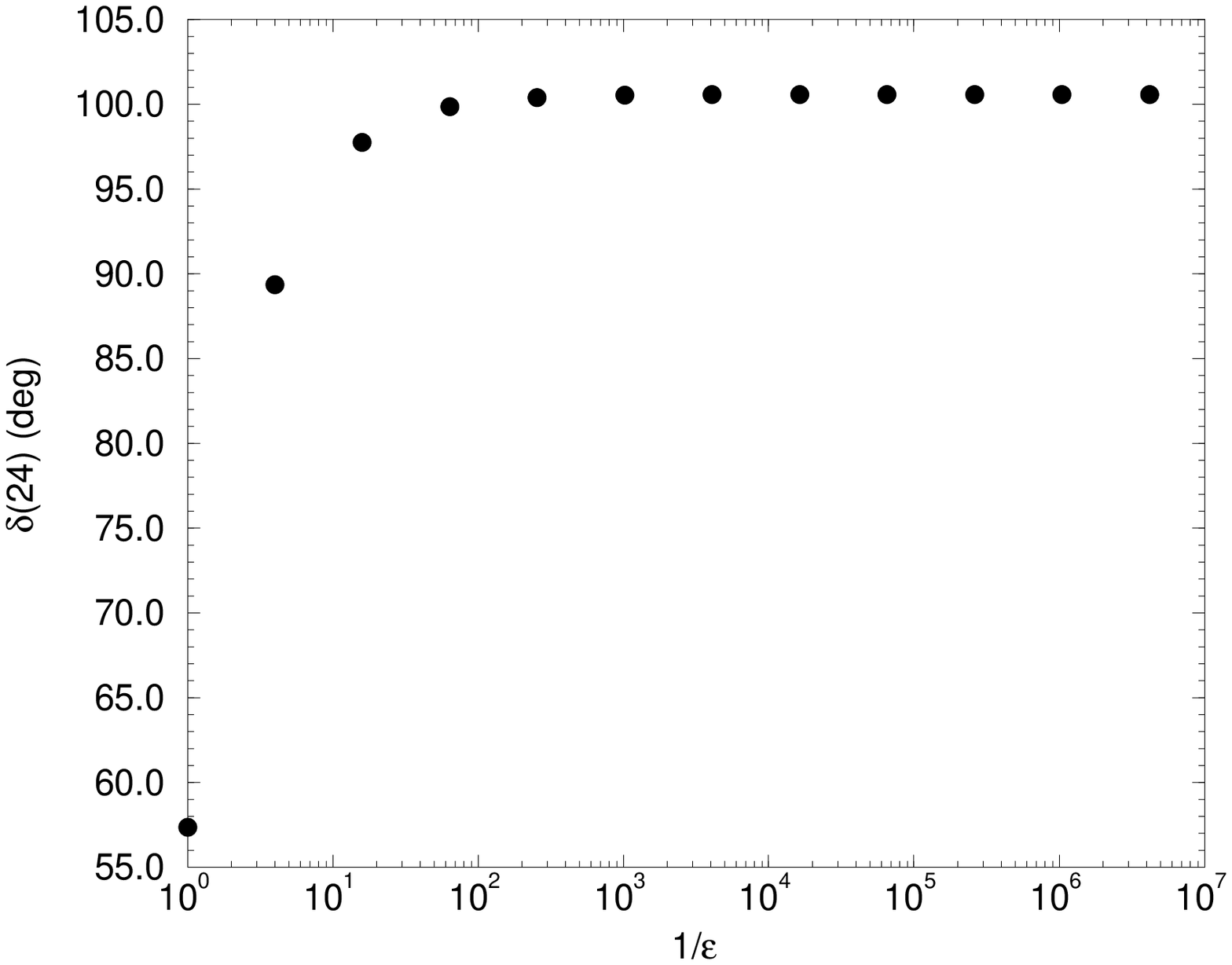,width=12.0cm}} 
\caption{Phase shifts at E$_{\rm Lab}=24.0$~MeV for the
solution of the LS equation as a function of $\epsilon$ for
$\epsilon\rightarrow 0$.}
\label{fig.2} 
\end{figure}

\begin{figure}[h]
\centerline{\epsfig{figure=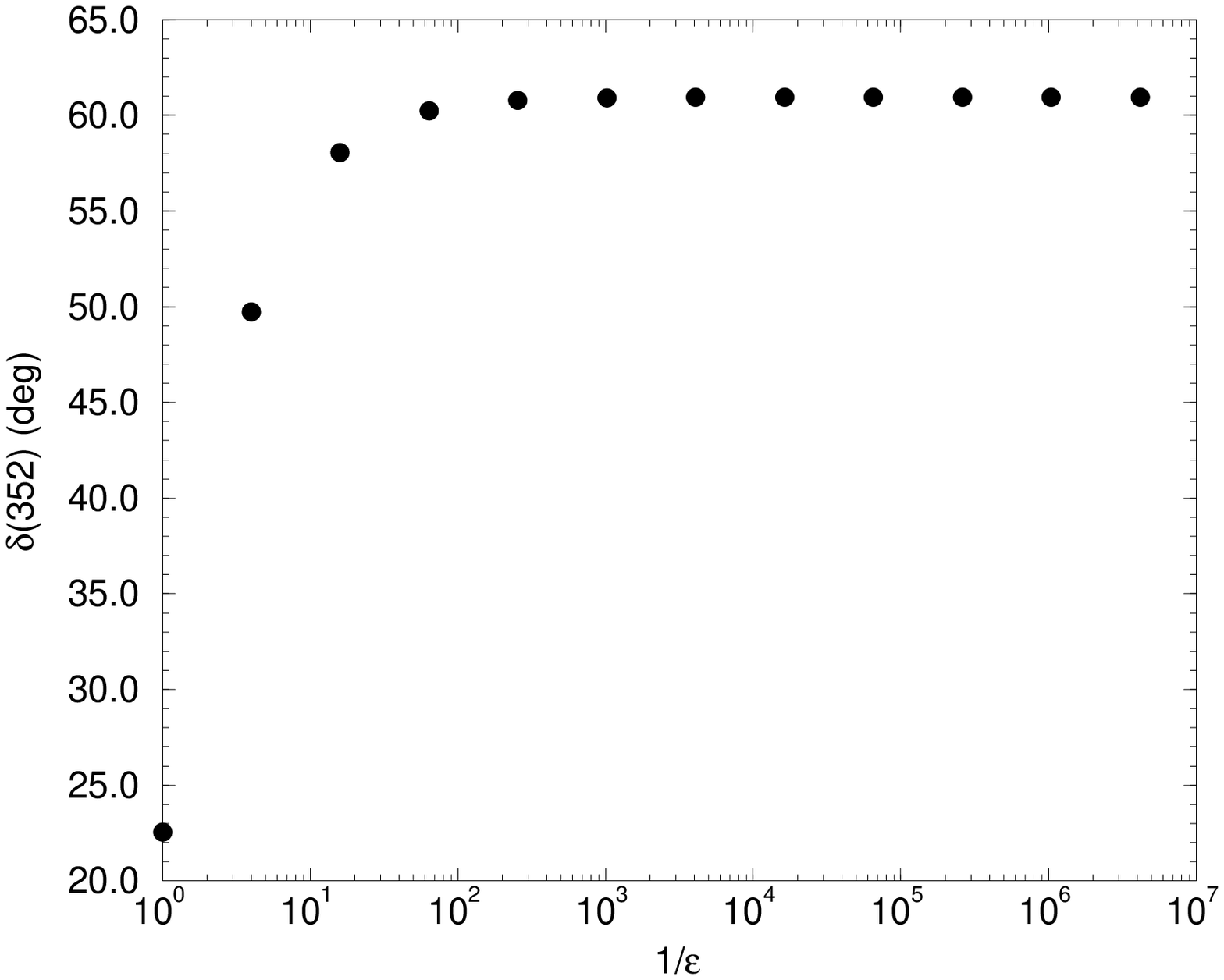, width=12.0cm}}
\caption{Phase shifts at E$_{\rm Lab}=352.0$~MeV
for the solution of the LS equation as
a function of $\epsilon$ for $\epsilon\rightarrow 0$.}\label{fig.3} 
\end{figure}

Second, we can examine the convergence of the phase shifts with
$\epsilon$. In Figs.~\ref{fig.2} and \ref{fig.3} we present the
results obtained from the Lippmann-Schwinger equation (\ref{eq:2.0.4})
with the potential (\ref{eq:2.1.1}) for the phase shifts as a function
of $1/\epsilon$.  In fact the points calculated using the two
different meshes described above are indistinguishable from one
another on the scale shown.  The calculation has been done for the
nucleon-nucleon system in the $^3$S$_1$ channel at laboratory energies
of 24 and 352 MeV.  The strength of the potential
$\lambda_\epsilon$ varies with $\epsilon$ as displayed in
Fig.~\ref{fig.lambda}, having been adjusted so that the potential
supports a bound state with a binding energy of 2.225 MeV.

A detailed comparison of the phase shifts resulting from the solution
of the dimensionally-regularized LS equation with those found using
the algebraic results of the previous section shows that the agreement
is good to four significant figures until we get to very small ($<
10^{-9}$) values of $\epsilon$. Long before that the phase shifts have
converged as a function of $\epsilon$. This comparison is presented in
in Tables~\ref{table-subtraction} and \ref{table-separable}, where the
the phase shifts obtained by these two methods at three energies, as
well as the bare coupling $\lambda_\epsilon$, are shown as a function
of $\epsilon$.  Table~\ref{table-subtraction} shows the results found
by solving the homogeneous, dimensionally-regularized LSE to get
$\lambda_\epsilon$ and then using that $\lambda_\epsilon$ to calculate
phase shifts in the dimensionally-regularized LSE~\footnote{The
results displayed in Table~\ref{table-subtraction} were obtained using
the K-matrix method and the second of the two meshes described above.}.
Table~\ref{table-separable} gives the result for phase shifts from
Eqs.~(\ref{eq:2.1.17}) and (\ref{eq:2.1.18}) as well as the result
(\ref{eq:2.1.13}) for $\lambda_\epsilon$.  We observe that there is
agreement between the algebraic and numerical renormalization to four
significant figures.  Furthermore, there is convergence in the phase
shift as $\epsilon\rightarrow 0$ to five significant figures, which is
beyond the numerical accuracy of this calculation. All this indicates
that the dimensionally-regularized integral equation is giving a
unique solution.

\begin{figure}[h]
\centerline{\epsfig{figure=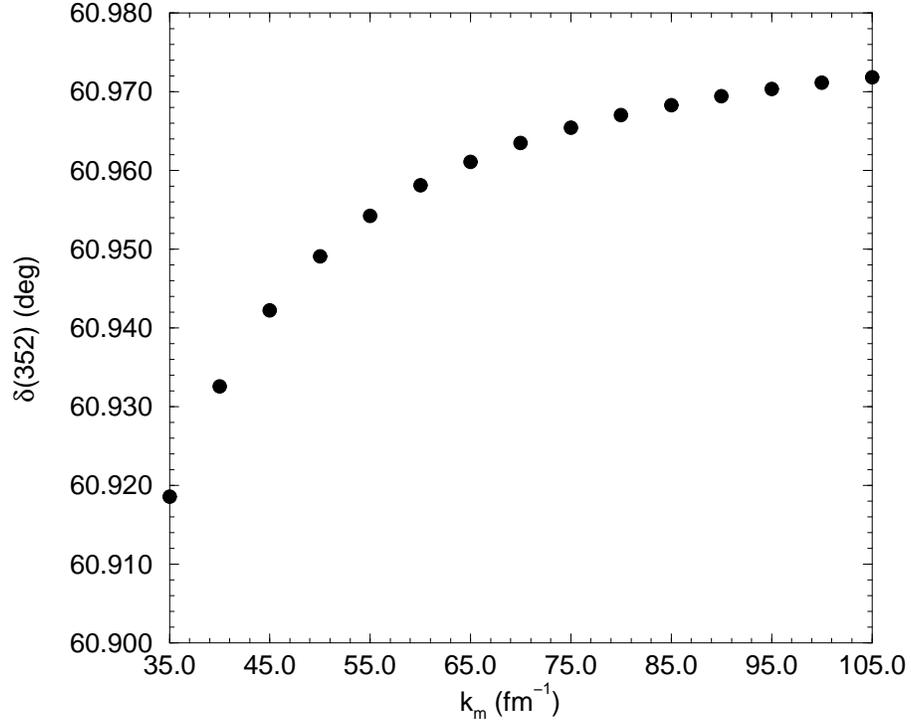, width=12.0cm}}
\caption{Phase shifts at E$_{\rm Lab}=352.0$~MeV with
$\epsilon=6.1035 \times 10^{-5}$, for the solution of the LS equation as
a function of $k_m$.}\label{fig-km}
\end{figure}

Finally, we should mention that there is some sensitivity to the value
of $k_m$ that is chosen. This is a numerical effect, and reflects the
wide spacing of quadratures in the logarithmic mesh above $k=k_m$. In
Figure~\ref{fig-km} we plot $\delta$ at $E_{\rm lab}=352$ MeV for
$\epsilon=6.1035 \times 10^{-5}$ over a range of $k_m$s, using meshes
of the second type described above. Provided $k_m$ is large enough the
results are stable to four significant figures. The variations in
other phase shifts, and in $\lambda_\epsilon$, are smaller than that
displayed in this plot.


\section{Conclusion}\label{sec5} 

From the above analysis and that of Ref.~\cite{ASW98} we conclude that
it is possible to use dimensional regularization to render divergent
integral equations with logarithmic divergences finite, provided that
a renormalization condition is imposed in order to fix one of the
parameters of the potential. Although in the present investigation we
have chosen the bound-state energy to fix the strength of the
potential, we could just as easily have used the scattering length or
the value of the on-shell amplitude at some finite energy as the
renormalization condition.

At this point one might ask whether the ideas discussed here can be
profitably employed if power-law divergences are involved. Of course,
such divergences do not appear explicitly when DR is implemented in
analytic calculations (see, e.g., Ref.~\cite{Ka96}). However, the
numerical techniques discussed above simply will not eliminate
ultra-violet divergences of degree greater than zero. The reason for
this lies in the way such divergences are eliminated in ``standard"
DR.  There the offending integral is analytically continued into a
region where $\epsilon$ is large enough so that the divergences no
longer appear. The resulting analytic form is then {\it defined} as
the value of the integral in the region where the integral was
formally divergent. It is this definition that eliminates the
power-law divergences, and, in contrast to the case where logarithmic
divergences are present, the limit $\epsilon \rightarrow 0$ cannot be
taken until this additional step is made. Since the work of this paper
relies on being able to straightforwardly take the limit $\epsilon
\rightarrow 0$ it is not clear that power-law divergences can be
eliminated in ``numerical" DR. It is possible that a numerical
procedure analogous to the analytic one just described can be
developed to make sense of integral equations in which power-law
divergences appear in the scattering series. However, the working out
of such a scheme is beyond the scope of this paper.

Although the present analysis is restricted to a simple toy model
which has only logarithmic divergences, we see no reason why it
cannot be easily extended to other integral equations of
interest. Two examples which we believe to be particularly important
are:

\begin{enumerate} 

\item Effective field theory treatments of neutron-deuteron scattering
in the doublet channel. The leading-order effective field theory
calculation produces Faddeev equations whose kernel does not go to
zero fast enough as $k \rightarrow \infty$ to make the integral
equation well-behaved. Normally this equation is regulated by a
cutoff, but the techniques discussed here could also be used. This
problem is particularly interesting because it has recently been shown
that unless a three-body force is added to the leading-order effective
field theory calculation the resultant amplitude is unduly
sensitive to the value of the cutoff~\cite{Be99}. This three-body
force introduces a new parameter into the calculation, which
is fit to the $NNN$ doublet scattering length, ${}^2a$. The
integral equation is then properly renormalized, in the sense that its
solutions are no longer sensitive to physics at short distances in the
$NNN$ system. However, formally it still contains divergences.

\item The Bethe-Salpeter equation for pion-nucleon
scattering~\cite{NT73,LA99}. In this case we would hope to use DR
rather than introduce form factors when solving the integral
equation. This might facilitate the introduction of electromagnetic
couplings into the $\pi-N$ scattering problem.

\end{enumerate}


\acknowledgments

We would like to thank Andreas Schreiber for stimulating our initial
interest in this study. We also thank Silas Beane for comments
on the manuscript. I.R.A. and A.G.H-E would like to thank the
Australian Research Council and Flinders University for their
support. D. R. P. thanks Flinders University and the Special Research
Structure for the Subatomic Structure of Matter for their warm
hospitality during the inception of this study. He is also grateful
to the U.~S. Department of Energy for its support under
contract no. DE-FG03-97ER4014.


\newpage

\begin{table}[h]
\begin{center}
\begin{tabular}{|c|c|c|c|c|}
  \hline $\epsilon$ & $\lambda$ (fm) & $\delta$(24) & $\delta$(96) &
  $\delta$(352) \\ \hline \hline 
  1.0 & 0.11928 & 57.352 & 37.151 & 22.553 \\ \hline

  2.5 $\times 10^{-1}$ & 4.9817 $\times 10^{-2}$ & 89.360 & 67.566 &
  49.740 \\ \hline 

  6.25 $\times 10^{-2}$ & 1.30340 $\times 10^{-2}$ & 97.748 & 76.120 &
  58.067 \\ \hline

  1.5625 $\times 10^{-2}$ & 3.27867 $\times 10^{-3}$ & 99.855 & 78.299 &
  60.224 \\ \hline

  3.90625 $\times 10^{-3}$ & 8.20662 $\times 10^{-4}$ & 100.38 & 78.848 &
  60.770 \\ \hline

  9.76563 $\times 10^{-4}$ & 2.05224 $\times 10^{-4}$ & 100.51 & 78.985 &
  60.906 \\ \hline
 
  2.44141 $\times 10^{-4}$ & 5.13094 $\times 10^{-5}$ & 100.55 & 79.019 &
  60.941 \\ \hline
 
  6.10352 $\times 10^{-5}$ & 1.28276 $\times 10^{-5}$ & 100.56 & 79.028 &
  60.949 \\ \hline
 
  1.52588 $\times 10^{-5}$ & 3.20691 $\times 10^{-6}$ & 100.56 & 79.030 &
  60.951 \\ \hline
 
  3.8147 $\times 10^{-6}$ & 8.01728 $\times 10^{-7}$ & 100.56 & 79.031 &
  60.952 \\ \hline

  9.53674 $\times 10^{-7}$ & 2.00432 $\times 10^{-7}$ & 100.56 & 79.031 &
60.952 \\ \hline
\end{tabular}
\end{center}
\caption{\label{table-subtraction}
  Couplings and phase shifts for various values of the distance
  $\epsilon$, away from three dimensions using the subtraction
  technique. All quantities are numerically accurate to four
  significant figures. Generated with $n_0=16$; $n_1=60$; $n_3=15$, 
   $k_{max}=50 \,
{\rm fm}^{-1}$.}
\end{table}

\begin{table}[h,t,b,p]
\begin{center}
\begin{tabular}{|c|c|c|c|c|}
  \hline $\epsilon$ & $\lambda$ (fm) & $\delta$(24) & $\delta$(96) &
  $\delta$(352) \\ \hline \hline 
  1.0 & 0.119286 & 57.356 & 37.154 & 22.555 \\ \hline

  2.5 $\times 10^{-1}$ & 4.98162 $\times 10^{-2}$ & 89.355 & 67.563 &
  49.738 \\ \hline 

  6.25 $\times 10^{-2}$ & 1.30340 $\times 10^{-2}$ & 97.742 & 76.114 &
  58.059 \\ \hline

  1.5625 $\times 10^{-2}$ & 3.27867 $\times 10^{-3}$ & 99.849 & 78.297 &
  60.222 \\ \hline

  3.90625 $\times 10^{-3}$ & 8.20662 $\times 10^{-4}$ & 100.38 & 78.845 &
  60.768 \\ \hline

  9.76563 $\times 10^{-4}$ & 2.05223 $\times 10^{-4}$ & 100.51 & 78.983 &
  60.905 \\ \hline
 
  2.44141 $\times 10^{-4}$ & 5.13094 $\times 10^{-5}$ & 100.54 & 79.017 &
  60.939 \\ \hline
 
  6.10352 $\times 10^{-5}$ & 1.28276 $\times 10^{-5}$ & 100.55 & 79.026 &
  60.948 \\ \hline
 
  1.52588 $\times 10^{-5}$ & 3.20691 $\times 10^{-6}$ & 100.55 & 79.028 &
  60.950 \\ \hline
 
  3.8147 $\times 10^{-6}$ & 8.01728 $\times 10^{-7}$ & 100.55 & 79.028 &
  60.950 \\ \hline

  9.5367 $\times 10^{-7}$ & 2.00432 $\times 10^{-7}$ & 100.55 & 79.028 &
  60.950 \\ \hline
\end{tabular}
\end{center}
\caption{\label{table-separable}
  Couplings and phase shifts for various values of the distance
  $\epsilon$, away from three dimensions using the algebraic solution
  of the separable potential and contour rotation. All quantities are
  numerically accurate to five significant figures. Generated with
  $n_0=16$, $n_1=80$, $n_2=10$, $n_3=15$, $\phi=0.7$ and $k_m=50$~fm$^{-1}$.}
\end{table}


\newpage

\end{document}